\documentclass[
    aps,
    prl,
    reprint,
    superscriptaddress,
    longbibliography
]{revtex4-2}

\usepackage{amsmath}
\usepackage{amssymb}
\usepackage{graphicx}
\usepackage{booktabs}
\usepackage{xcolor}

\usepackage[
    colorlinks=true,
    citecolor=blue!55!black,
    linkcolor=blue!55!black,
    urlcolor=blue!55!black
]{hyperref}

\usepackage{orcidlink}

\hypersetup{
    pdftitle={The Schwarzschild Ringdown Time Prescription on Critical Kerr Plunges},
    pdfauthor={Jingxu Wu, Junyi Zhang, Daniil Stepanenko, Maxim Khlopov}
}

\newcommand{\head}[1]{\textit{#1.}\enspace}
\newcommand{\dd}{\mathrm{d}}

\begin{document}

\title{
    The Schwarzschild Ringdown Time Prescription
    on Critical Kerr Plunges
}

\author{Jingxu Wu\,\orcidlink{0009-0003-5685-8188}}
\email{wuxj@my.msu.ru}
\affiliation{
    Faculty of Physics,
    Lomonosov Moscow State University,
    Moscow 119991, Russia
}

\author{Junyi Zhang}
\email{st02240136@gse.cs.msu.ru}
\affiliation{
    Faculty of Computational Mathematics and Cybernetics,
    Lomonosov Moscow State University,
    Moscow 119991, Russia
}

\author{Daniil Stepanenko\,\orcidlink{0000-0002-9771-6632}}
\email{dstepanenko@mi-ras.ru}
\affiliation{
    Steklov Mathematical Institute,
    Russian Academy of Sciences,
    Moscow 119333, Russia
}

\author{Maxim Khlopov\,\orcidlink{0000-0002-1653-6964}}
\email{maxim.khlopov@cern.ch}
\affiliation{
    Virtual Institute of Astroparticle physics,
    75018, Paris, France
}
\begin{abstract}
For a Schwarzschild plunge, constant-amplitude quasinormal-mode convergence has been tied to the source's advanced time at $r=0$, lying $9.435M$ after the $(2,2)$ strain peak. On real prograde critical Kerr plunges, $P(r_-)<0$ forces the ingoing time $v$ to diverge logarithmically at the inner horizon while $u_K$ stays finite. Matching $v$ to $-u_K$ at the unique interior zero $P(r_0)=0$ yields a parameter-free onset hierarchy $(u_{\rm H},u_{r_0},u_{\rm C})$ that recovers $9.435M$ as $a\to 0$ and predicts two-stage residual decay and frequency locking at $(6.85M,11.35M,13.35M)$ on an $a/M=0.9$ eccentric plunge.
\end{abstract}
\maketitle

\begin{figure*}[t]
\centering
\includegraphics[width=.94\textwidth]{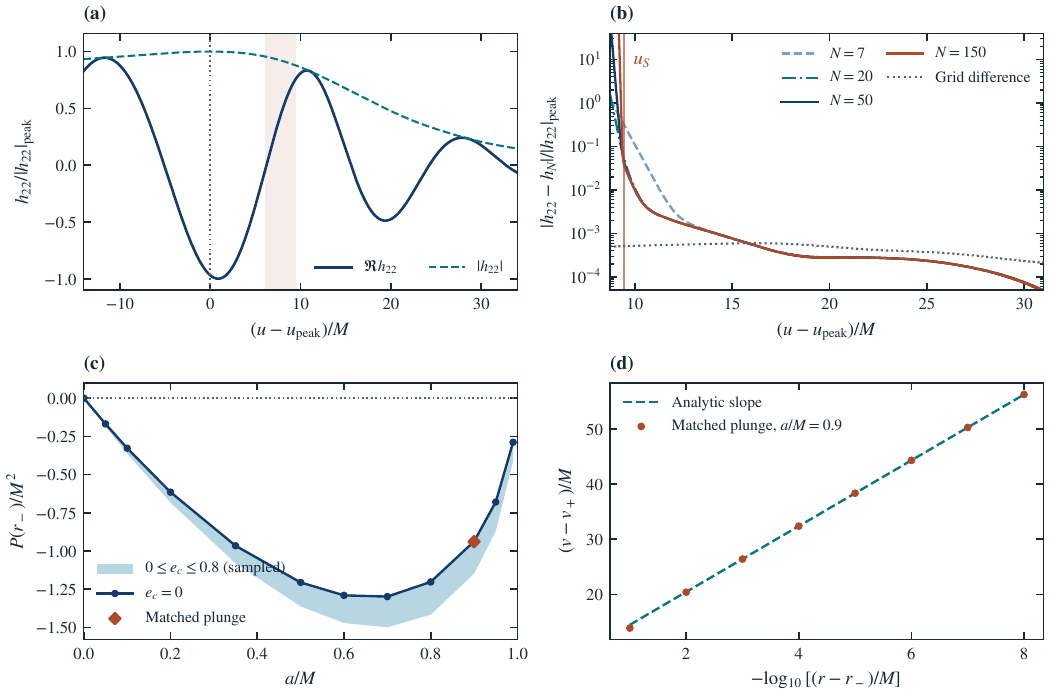}
\caption{Schwarzschild waveform benchmark and Kerr branch test. (a) Independently evolved Schwarzschild Zerilli $(2,2)$ strain normalized by its peak; the shaded interval lies between the tortoise bounce $u_B$ and $u_S$. (b) Relative residuals of prescribed constant-amplitude QNM sums through overtone index $N$; the dotted line shows the difference between the two radial grids. (c) Inner-horizon canonical momentum $P(r_-)/M^2$ across spin; the shaded band spans critical eccentricities $0\leq e_c\leq 0.8$ and the diamond marks the $a/M=0.9$ FEW-matched plunge. (d) Logarithmic divergence of the ingoing Kerr time $v-v_+$ approaching $r_-$ on the matched plunge versus the analytic slope of Eq.~\eqref{eq:log}.}
\label{fig:main}
\end{figure*}

\head{Introduction}
A perturbed black hole relaxes toward equilibrium by radiating gravitational waves whose late-time spectrum is governed by discrete, complex quasinormal-mode (QNM) frequencies~\cite{regge,zerilli,vishveshwara,teukolsky,leaver85,kokkotas,berti09}. Because the QNM spectrum of a Kerr remnant in general relativity depends solely on its mass and spin, measuring multiple ringdown modes provides a direct test of the no-hair theorem and strong-field gravity~\cite{isi,cotesta,franchini,carullo25,barack19,gair13}. A central theoretical and observational question in black-hole spectroscopy is when an exterior waveform ceases to be dominated by transient source driving and enters the regime where a sum of damped sinusoids with constant amplitudes is physically meaningful~\cite{blackman15,blackman17,bernard19}. Empirical fits starting near the peak of the $(2,2)$ strain amplitude can achieve small waveform residuals when seven or more overtones are included as free parameters~\cite{giesler,isi}, yet agnostic spectroscopy analyses and controlled linear-perturbation studies demonstrate that fitted amplitudes near the peak vary strongly with the fitting window and absorb prompt response, spherical--spheroidal mode mixing, power-law tails, and second-order nonlinearities~\cite{cotesta,baibhav,cheung,mitman,lagos23,qqnm25}.

Separating empirical window fitting from the mathematical convergence of a prescribed constant-amplitude QNM expansion requires analyzing the retarded Green function in the complex frequency plane~\cite{leaver86,zimmerman16,mark17,hyperboloidal25}. In Leaver's spectral decomposition~\cite{leaver86}, closing the Fourier contour in the lower half-plane extracts the discrete QNM residue series only after the high-frequency semicircle and source integral become exponentially suppressed at large overtone index $N$. Recently, Kuntz~\cite{kuntz,kuntzdata} established a sharp convergence boundary $u_S$ for a point particle plunging into a Schwarzschild black hole by analytically continuing the distributional source along the real inward trajectory to the central singularity $r=0$. Because the ingoing Eddington--Finkelstein advanced time $v=t+x$ of a Schwarzschild plunge remains finite at $r=0$, the high-overtone asymptotic phase closes at $u_S=v(0)$, which follows the $(2,2)$ strain peak by approximately $9.435M$ for a plunge from the innermost stable circular orbit (ISCO)~\cite{kuntz}.

Extending this prescription to a rotating Kerr black hole is vital for astrophysical coalescences and extreme-mass-ratio inspirals (EMRIs), yet it confronts fundamental geometric and wave-equation differences. In Kerr spacetime, gravitational perturbations are governed by the spin-weight $s=-2$ Teukolsky equation and its Sasaki--Nakamura transformation~\cite{teukolsky,sasaki,pound21,stein19}, where frame dragging couples the radial barrier to the outer-horizon angular velocity $\Omega_+=a/(2Mr_+)$ and splits the horizon into an outer event horizon $r_+$ and an inner Cauchy horizon $r_-$. Recent frequency- and time-domain studies of particles plunging into Kerr have examined how the transition from source-driven oscillations to free ringdown depends on the photon-ring barrier and horizon redshift~\cite{weller,kubota}, while point-particle Teukolsky evolutions show that orbital eccentricity, spin alignment, and the radial phase at merger leave pronounced imprints on the late waveform and QNM excitation factors~\cite{babak07,sundararajan07,sundararajan10,harms,taracchini,apte,faggioli,beckerwave}.

At the orbital level, a compact body driven by radiation reaction evolves from an adiabatic inspiral through a boundary-layer transition near the separatrix into a geodesic plunge~\cite{orithorne,compere,beckertransition,few,lisawg}. On the inward geodesic plunge, the radial motion is integrable in Mino time and admits exact factorizations across spin and eccentricity~\cite{mummery,dyson}. However, whether the orbital advanced time $v=t+r_*$ used in the Schwarzschild ringdown prescription remains finite along a real Kerr plunge—and how the interior horizon structure constrains exterior ringdown convergence—has remained open.

In this Letter, we resolve this question in two stages. First, we prove that on every prograde bound critical equatorial Kerr plunge ($0<a<M$, $0\leq e_c<1$), the canonical momentum combination $P(r)=E(r^2+a^2)-aL_z$ is strictly negative at the inner horizon $r_-$. Consequently, the ingoing Kerr time $v$ diverges logarithmically as $r\to r_-$, whereas the complementary outgoing coordinate $u_K=t-r_*$ remains finite, creating a coordinate-branch obstruction to applying the unmodified Schwarzschild prescription. We confirm this branch sign and logarithmic slope across a 96-orbit critical grid and on an $a/M=0.9$ eccentric plunge matched to a FastEMRIWaveforms (FEW) inspiral~\cite{few,fewsoftware}. Second, we show that because $\dd P/\dd r=2Er>0$ with $P(r_+)>0$ and $P(r_-)<0$, every prograde critical plunge possesses a unique interior frame-dragging reversal radius $r_0=\sqrt{aL_z/E-a^2}\in(r_-,r_+)$ where $P(r_0)=0$. Matching the outer-horizon-regular clock $v$ on $[r_0,r_{\rm lr}]$ to the inner-horizon-regular clock $-u_K$ on $[r_-,r_0]$ yields a parameter-free, three-stage exterior onset hierarchy $(u_{\rm H},u_{r_0},u_{\rm C})$ that smoothly recovers the Schwarzschild benchmark as $a\to 0$ and produces falsifiable predictions for overtone convergence, frequency locking, and eccentricity shifts verified against sourced Kerr evolutions and public Teukolsky impulse data~\cite{kubota,impulsedata,dali,vwc}.

\head{Schwarzschild benchmark}
Throughout we set $G=c=1$, and $M$ denotes the background mass. Write $x=r+2M\ln(r/(2M)-1)$ for the Schwarzschild tortoise coordinate. Set $t=0$ at the light-ring crossing. The bounce occurs at $x=0$. With $\rho=r/M$, $\ell=L_z/M$ and specific energy $E$, the timelike radial equation has
\begin{equation}
\mathcal R(\rho)=E^2-(1-2/\rho)(1+\ell^2/\rho^2).
\end{equation}
The start-to-bounce interval of Ref.~\cite{kuntz} is
\begin{equation}
\frac{u_S-u_B}{M}=
\int_0^{\rho_B}\!
\frac{1+\ell^2/\rho^2}{\sqrt{\mathcal R}\left(E+\sqrt{\mathcal R}\right)}\,
\dd\rho .
\label{eq:schwarz}
\end{equation}
For the ISCO plunge, $E=\sqrt{8/9}$, $\ell=\sqrt{12}$ and $\rho_B=2.556929$. Equation~\eqref{eq:schwarz} gives $(u_S-u_B)/M=3.372244$ and $u_S/M=6.999763$ in this normalization.

We evolve the even-parity Zerilli equation~\cite{zerilli} with the moving distributional source of Ref.~\cite{kuntz}. The source is switched on over $-600M<t<-400M$. The evolution uses a compactified radial grid and fourth-order Runge--Kutta integration, representing the point source and its derivative with narrow Gaussians~\cite{sundararajan07,sundararajan10,harms}. The two radial grid spacings are $0.05M$ and $0.025M$. At future null infinity the strain convention is $rh_{22}/\mu=\sqrt{24}\,\Psi_{22}$. The fine-grid peak at $u_{\rm peak}/M=-2.435$ gives $(u_S-u_{\rm peak})/M=9.435$, in agreement with the reported Schwarzschild value~\cite{kuntz}.

Figures~\ref{fig:main}(a,b) compare the waveform with the behavior of the fixed-amplitude residual. We use the published coefficients~\cite{kuntzdata}, including the supplied algebraically special contribution~\cite{leaver85,kokkotas}. At $u=u_S+2M$ the $N=150$ residual is $2.42\times10^{-3}$ of the peak, and the two numerical solutions differ by $5.43\times10^{-4}$ of the peak. The high overtones increase the residual before $u_S$. The residual also contains the branch-cut contribution~\cite{leaver86,hyperboloidal25}, which we leave unevaluated in this calculation.

\head{Critical Kerr plunges}
Let $0<a<M$ and consider an equatorial orbit. Define $B=L_z-aE$ and
\begin{align}
r_\pm &= M\pm\sqrt{M^2-a^2},\\
\Delta &= (r-r_+)(r-r_-),\\
P(r) &= E(r^2+a^2)-aL_z.
\end{align}
With Mino time $\dd\lambda=\dd\tau/r^2$ and proper time $\tau$, the inward branch obeys~\cite{dyson,mummery}
\begin{equation}
\frac{\dd r}{\dd\lambda}=-\sqrt{R},\qquad
R=P^2-\Delta(r^2+B^2).
\label{eq:R}
\end{equation}
For a bound critical plunge, the constants are those of the equatorial separatrix orbit~\cite{mummery,dyson,apte}. If $r_u$ is the repeated unstable circular radius and $r_a$ the outer turning radius, then
\begin{equation}
R=(1-E^2)r(r_a-r)(r_u-r)^2 .
\label{eq:factor}
\end{equation}
The circular ISCO has $r_a=r_u$. At fixed critical eccentricity $0\leq e_c<1$, $E^2<1$ and $r_a\geq r_u>r_+>r_->0$. Hence $P(r_-)^2=R(r_-)>0$. At small positive spin, $P(r_-)=2Mr_-E-aL_z<0$ because $r_-=O(a^2/M)$ and $L_z>0$ on the prograde branch. The branch is continuous and Eq.~\eqref{eq:factor} forbids a zero crossing. This fixes the sign throughout the branch.

Define $v=t+r_*$ and the complementary Kerr coordinate $u_K=t-r_*$, with $\dd r_*/\dd r=(r^2+a^2)/\Delta$. Their evolution along the inward branch is
\begin{align}
\frac{\dd v}{\dd\lambda}
  &= aB+(r^2+a^2)\frac{P-\sqrt R}{\Delta},\label{eq:v_rate}\\
\frac{\dd u_K}{\dd\lambda}
  &= aB+(r^2+a^2)\frac{P+\sqrt R}{\Delta}.
\label{eq:pair}
\end{align}
A crossing orbit has $P(r_+)>0$ at the future outer horizon, where the apparent pole in $\dd v/\dd\lambda$ cancels. At the inner horizon, $P(r_-)<0$ gives $P-\sqrt R\to2P(r_-)$. Dividing by $\dd r/\dd\lambda$ and integrating from $r>r_-$ yields
\begin{equation}
\frac{v}{M}=-\frac{4r_-}{r_+-r_-}
 \ln\!\left(\frac{r-r_-}{M}\right)+O(1).
\label{eq:log}
\end{equation}
The complementary numerator in Eq.~\eqref{eq:pair} vanishes on that branch. In particular, $(P+\sqrt R)/\Delta=(r^2+B^2)/(P-\sqrt R)$ is finite at $r_-$, as is the proper time remaining to that radius. The divergence selects a coordinate branch. As $a\to0$, the coefficient tends to zero and the inner horizon shrinks.

The critical-orbit grid contains 96 configurations at 12 spins from $0$ to $0.99$ and eight eccentricities from $0$ to $0.8$. Each $r_u=p_{\rm sep}/(1+e_c)$ determines the circular-orbit constants. All 88 rotating configurations have $P(r_-)<0$ Fig.~\ref{fig:main}(c). Numerical integration of $\dd v/\dd r$ over successive logarithmic intervals recovers the analytic coefficient with a maximum relative discrepancy of $1.4\times10^{-7}$ over the final tested intervals. At $a=0$ and $e_c=0$, the advanced-time interval from the light ring to the center is $5.38606M$, which recovers the Schwarzschild reference.

\head{Matched eccentric example}
We connect a FastEMRIWaveforms (FEW) trajectory~\cite{few,fewsoftware,lisawg} to an approximate radiation-reaction transition and a geodesic plunge~\cite{orithorne,compere,beckertransition}. The baseline has $M=10^6M_\odot$, $\mu=10M_\odot$, $a/M=0.9$, $p_0/M=8.5$ and $e_0=0.3$. FEW reaches its buffer $p-p_{\rm sep}=0.002M$ with $e=0.06168$, after a minimum near $0.05005$. The adiabatic $(2,2)$ waveform spans $724.85$ days at this mass scale and includes the radial harmonics $n=-55,\ldots,55$.

At the match we impose continuity of radius and its coordinate-time derivative. We continue a local torque and taper it with radius, with $\dd E=\Omega_{\phi,\mathrm{inst}}\dd L_z$. Together with the radial acceleration $\partial_rR/2$, this preserves $(\dd r/\dd\lambda)^2-R$ in the transition equations. The instantaneous radial-speed match shifts $E$ by about $-9.95\times10^{-6}$ from its adiabatic value. Below the switch radius we freeze $E$ and $L_z$. The local torque approximation differs from the transition prescription of Becker and Hughes~\cite{beckertransition}.

\begin{figure}[t]
\centering
\includegraphics[width=\columnwidth]{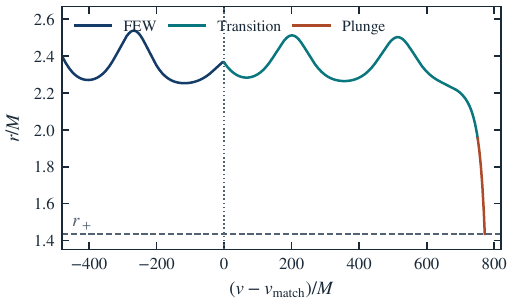}
\caption{Radial motion of the $a/M=0.9$ baseline through the FEW match, transition and geodesic plunge. The dotted line fixes the match origin. The dashed line marks the outer horizon.}
\label{fig:orbit}
\end{figure}
\begin{figure*}[!t]
\centering
\includegraphics[width=.94\textwidth]{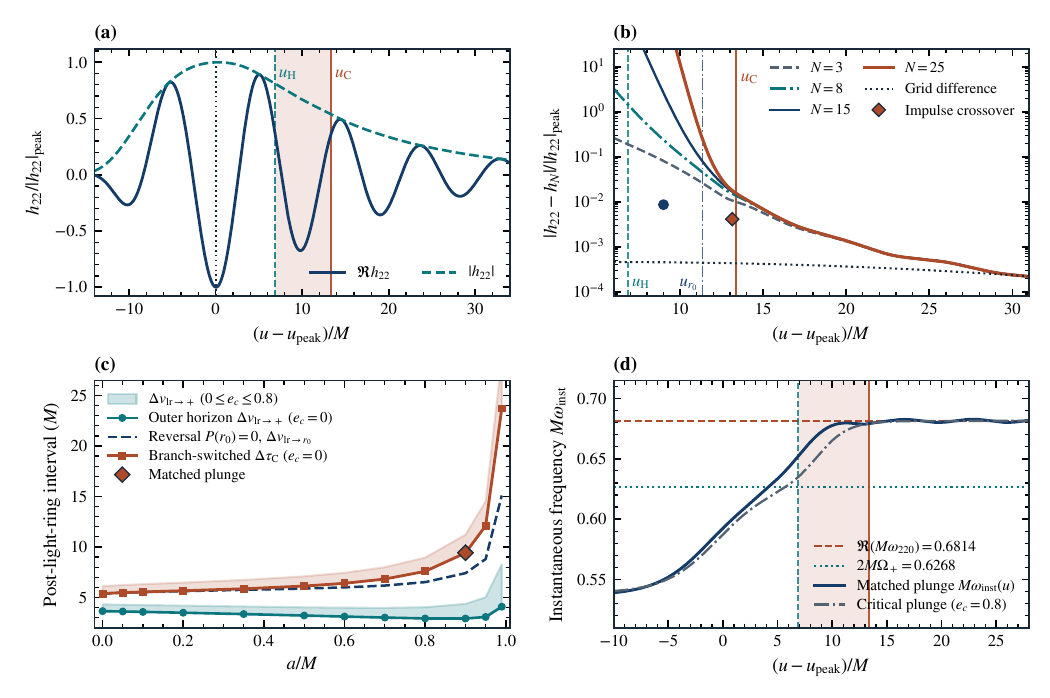}
\caption{Testable exterior Kerr waveform predictions and verification on the $a/M=0.9$ matched eccentric plunge. (a) Sourced exterior $(2,2)$ strain normalized by its peak, with the transition window shaded between the outer-horizon causal boundary $u_{\rm H}$ and the branch-switched Cauchy boundary $u_{\rm C}$. (b) Relative residuals of prescribed constant-amplitude Kerr QNM sums through prograde overtone index $N$ (plus two retrograde modes), crossing the reversal boundary $u_{r_0}$ and converging at $u_{\rm C}$; the circle and diamond mark the $1\%$ passing start ($9.0M$) and the $2+2$/$4+2$ overtone crossover ($13.14M$) in the independent public Teukolsky impulse data~\cite{kubota,impulsedata}. (c) Post-light-ring onset hierarchy $\Delta v_{\mathrm{lr}\to +}$, $\Delta v_{\mathrm{lr}\to r_0}$, and $\Delta\tau_{\rm C}$ across spin and critical eccentricity $0\leq e_c\leq 0.8$. (d) Instantaneous exterior frequency $M\omega_{\rm inst}(u)$ rising through the horizon frame-dragging rate $2M\Omega_+=0.6268$ near $u_{\rm H}$ and locking onto $\Re(M\omega_{220})=0.6814$ at $u_{\rm C}$.}
\label{fig:prediction}
\end{figure*}

The geodesic segment in Fig.~\ref{fig:orbit} has $E=0.8447723$ and $L_z/M=2.1020428$. Its continuation gives $P(r_-)/M^2=-0.93875$ and a fitted coefficient of $2.58832$ in Eq.~\eqref{eq:log} Fig.~\ref{fig:main}(d). Changing the three initial matching buffers gives intervals of $405M$ to $1218M$ between match and outer horizon. At fixed initial match, changing the geodesic switch alters that interval by less than $0.02M$ for the tested settings. Across all five runs, the photon-orbit-to-outer-horizon interval is $2.922M$ to $2.929M$. These advanced-time measurements all refer to the orbit.

\head{Testable exterior Kerr waveform predictions}
Because $\dd P/\dd r=2Er>0$ for $r>0$ with $P(r_+)>0$ and $P(r_-)<0$, every prograde critical plunge possesses a unique interior frame-dragging reversal radius
\begin{equation}
r_0 = \sqrt{\frac{aL_z}{E}-a^2}\in(r_-,r_+),\qquad P(r_0)=0,
\label{eq:r0}
\end{equation}
where the Boyer--Lindquist coordinate-time rate $(r^2+a^2)P/\Delta$ changes sign. Rewriting Eqs.~\eqref{eq:v_rate} and~\eqref{eq:pair} in pole-factored form,
\begin{align}
-\frac{\dd v}{\dd r} &= \frac{1}{\sqrt R}\left[aB+\frac{(r^2+a^2)(r^2+B^2)}{P+\sqrt R}\right], \label{eq:dv_reg}\\
-\frac{\dd(-u_K)}{\dd r} &= \frac{1}{\sqrt R}\left[-aB+\frac{(r^2+a^2)(r^2+B^2)}{|P|+\sqrt R}\right], \label{eq:duK_reg}
\end{align}
shows that $v(r)$ is strictly analytic on $[r_0,r_{\rm lr}]$ ($P\geq 0$), whereas $-u_K(r)$ is strictly analytic on $[r_-,r_0]$ ($P\leq 0$), with their horizon fractions matching continuously at $r_0$ to $\sqrt{-(r_0^2+B^2)/\Delta(r_0)}$. Switching from the outer-horizon-regular clock $v$ to the inner-horizon-regular clock $-u_K$ at $r_0$ defines a parameter-free post-light-ring continuation interval
\begin{equation}
\Delta\tau_{\rm C}(a,e_c) \equiv \Delta v_{\mathrm{lr}\to +} + \Delta v_{+\to r_0} + \Delta u_{K,r_0\to -},
\label{eq:tauC}
\end{equation}
where $\Delta v_{\mathrm{lr}\to +}=v(r_+)-v(r_{\rm lr})$, $\Delta v_{+\to r_0}=v(r_0)-v(r_+)$, and $\Delta u_{K,r_0\to -}=u_K(r_0)-u_K(r_-)$ are invariant under additive shifts of $r_*$. As $a\to 0$, both $r_-=O(a^2/M)$ and $r_0=O(\sqrt{aM})$ collapse to $r=0$, so $\Delta u_{K,r_0\to -}\to 0$ and $\Delta\tau_{\rm C}(0,0)=\Delta v_{\mathrm{lr}\to r_0}(0,0)=5.38606M$, recovering the exact Schwarzschild light-ring-to-center interval and Eq.~\eqref{eq:schwarz} without arbitrary cutoffs.

Together with the exterior barrier-reflection offset $\Delta u_{\rm refl}(a,e_c)=[u_{\rm lr}-u_{\rm peak}]+2[r_*(r_{\rm lr})-r_*(r_B)]$ ($4.049M$ at $a=0$ and $3.929M$ at $a/M=0.9$), the three radii $(r_+,r_0,r_-)$ yield a falsifiable three-stage hierarchy of exterior times relative to the $(2,2)$ strain peak:
\begin{equation}
u_{\rm H}-u_{\rm peak} < u_{r_0}-u_{\rm peak} < u_{\rm C}-u_{\rm peak},
\label{eq:hierarchy}
\end{equation}
obtained by adding $\Delta v_{\mathrm{lr}\to +}$, $\Delta v_{\mathrm{lr}\to r_0}=\Delta v_{\mathrm{lr}\to +}+\Delta v_{+\to r_0}$, and $\Delta\tau_{\rm C}$ to $\Delta u_{\rm refl}$, respectively Fig.~\ref{fig:prediction}(c). On the $a/M=0.9$ FEW-matched eccentric plunge ($r_0=1.19560M$), Eq.~\eqref{eq:tauC} predicts $\Delta v_{\mathrm{lr}\to +}=2.92536M$, $\Delta v_{\mathrm{lr}\to r_0}=7.41958M$, and $\Delta\tau_{\rm C}=9.42534M$, placing the three exterior boundaries at $(u_{\rm H}-u_{\rm peak})/M=6.854$, $(u_{r_0}-u_{\rm peak})/M=11.348$, and $(u_{\rm C}-u_{\rm peak})/M=13.354$.

This geometric structure yields three concrete predictions for exterior retarded spin-$-2$ waveforms~\cite{harms,beckerwave,kubota,weller}, tested in Fig.~\ref{fig:prediction}:
(i) \textit{Two-stage overtone convergence:} For $u<u_{\rm H}$, the source has not cleared the outer-horizon redshift zone ($2M\Omega_+=a/r_+=0.62679$, surface gravity $M\kappa_+=0.15178$), so fixed-amplitude overtone sums diverge rapidly with overtone index $N$. Between $u_{\rm H}$ and $u_{r_0}$, the horizon-driven tail decays as $\exp[-2\kappa_+(u-u_{\rm H})]$, while full multi-overtone convergence sets in across the branch-switched boundary $u_{\rm C}$ Fig.~\ref{fig:prediction}(a,b), where the $N=25$ residual drops to $5.77\times 10^{-3}$ of the peak at $u_{\rm C}+2M$. In the independent public Teukolsky impulse-response series at $a/M=0.9$~\cite{kubota,impulsedata}, the $2+2$ mode holdout error first passes the $1\%$ threshold at $9.0M\in[u_{\rm H}-u_{\rm peak},u_{r_0}-u_{\rm peak}]$ ($8.64\times 10^{-3}$) and crosses the $4+2$ mode curve at $13.14M$ ($4.12\times 10^{-3}$), within $0.21M$ of $u_{\rm C}-u_{\rm peak}=13.354M$ Fig.~\ref{fig:prediction}(b), while at $a/M=0.99$ the strain onset shifts to $13M$ as $\Delta v_{\mathrm{lr}\to r_0}$ expands to $15.10M$.
(ii) \textit{Driven-to-free frequency locking:} The instantaneous exterior frequency $M\omega_{\rm inst}(u)=-\mathrm{Im}(\dot h_{22}/h_{22})$ crosses the horizon frame-dragging rate $2M\Omega_+=0.6268$ near $u_{\rm H}$ and locks onto the fundamental Kerr QNM frequency $\Re(M\omega_{220})=0.68143$ at $u_{\rm C}$ Fig.~\ref{fig:prediction}(d).
(iii) \textit{Transition-buffer invariance versus separatrix eccentricity shift:} Across all five FEW-to-plunge transition runs, despite an $813M$ spread in match-to-horizon duration ($405M$ to $1218M$), the branch-switched interval $\Delta\tau_{\rm C}$ varies by only $0.0077M$ ($9.4218M$ to $9.4295M$), predicting universal post-peak waveforms with mismatch $\mathcal M_{[u_{\rm peak},u_{\rm peak}+35M]}<4\times 10^{-5}$. Conversely, varying the critical eccentricity $0\leq e_c\leq 0.8$ at $a/M=0.9$ shifts $\Delta\tau_{\rm C}$ by $+1.812M$ ($9.410M$ to $11.222M$), over $200$ times larger than the transition-prescription spread Fig.~\ref{fig:prediction}(c,d).

\head{Data availability}
The accompanying replication files contain the coordinate trajectories, the Schwarzschild residuals, the Kerr benchmark tables, and the exterior waveform prediction code. The Supplemental Material identifies the public waveform and coefficient inputs.

\head{Acknowledgments}
The~research of M.K. was carried out in the Southern Federal University with financial support from the Ministry of Science and Higher Education of the Russian Federation (State contract FENW-2026-0028).

\bibliography{references}
\clearpage
\onecolumngrid
\setcounter{figure}{0}
\setcounter{table}{0}
\renewcommand{\thefigure}{S\arabic{figure}}
\renewcommand{\thetable}{S\arabic{table}}
\renewcommand{\theHfigure}{supp.\arabic{figure}}
\renewcommand{\theHtable}{supp.\arabic{table}}
\section*{Supplemental Material}
\noindent\textit{Position of the waveform tests.}
The main text establishes the interior frame-dragging reversal radius $r_0=\sqrt{aL_z/E-a^2}\in(r_-,r_+)$ and the resulting three-stage exterior onset hierarchy $(u_{\rm H},u_{r_0},u_{\rm C})$. The following waveform tests check the finite-mode fitting procedure, the benchmark spin-eccentricity tables, and the behavior of independent comparison data. The impulse calculations use publicly archived Teukolsky time series~\cite{kubota,impulsedata}. The 22 series contain $h_{22}$ and $\psi_{4,22}$ over 11 spins from zero to $0.99$. Each retains its published field convention. Its amplitude peak fixes the time origin.

For a trial start $u_0$, we fit complex amplitudes to fixed Kerr frequencies on $[u_0,u_0+24M)$. We use $N\in\{2,4,6\}$ prograde modes and two retrograde modes. The relative holdout error is the complex residual norm divided by the data norm on $[u_0+24M,u_0+44M)$. We transport the fitted fundamental amplitude to the common time origin for comparisons. A setting passes when the holdout error is below $1\%$ before and after factor-two decimation and the fundamental amplitude changes by less than $5\%$ under a $1M$ start shift and under decimation. The criterion must hold over three consecutive starts on the $1M$ grid. Decimation tests the fit's sensitivity to the samples in the published waveform.

\begin{figure}[!htbp]
\includegraphics[width=.76\linewidth]{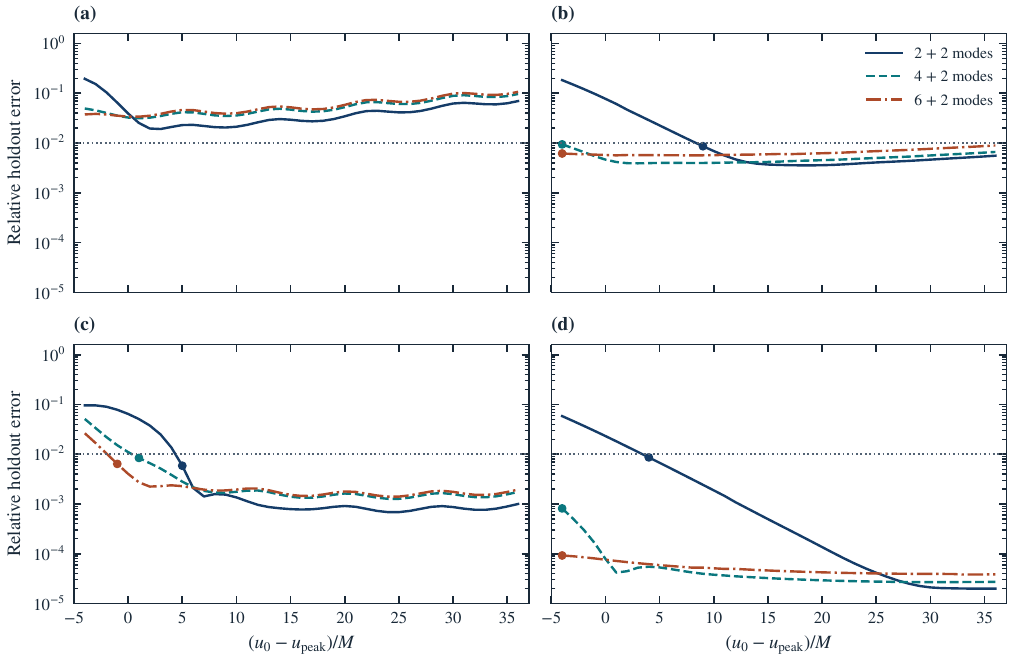}
\caption{\label{fig:impulse-validation}Relative holdout error for four impulse-response cases: (a) $h_{22}$ at $a/M=0$, (b) $h_{22}$ at $a/M=0.9$, (c) $\psi_{4,22}$ at $a/M=0$, and (d) $\psi_{4,22}$ at $a/M=0.9$. The dotted line shows the $1\%$ threshold. A point marks the first start that satisfies all error and stability tests for three consecutive starts. Each field uses its own peak as the time origin.}
\end{figure}

\smallskip
We evaluated 2706 fit configurations across the start range $[-4M,36M]$. Of the 66 field, spin and mode-count settings, 41 contain a persistent passing interval. For four prograde and two retrograde modes at $a/M=0.9$, both fields pass at the earliest sampled start. For two prograde and two retrograde modes at $a/M=0.9$ [Fig.~\ref{fig:impulse-validation}(b)], the strain holdout error drops from $1.09\times 10^{-2}$ at $8M$ to $8.64\times 10^{-3}$ at $(u_0-u_{\rm peak})/M=9.0$—inside the predicted exterior horizon-to-reversal interval $[6.854,11.348]$—and crosses the $4+2$ mode curve at $(u_0-u_{\rm peak})/M=13.14$ ($4.12\times 10^{-3}$), within $0.21M$ of the branch-switched Cauchy boundary $(u_{\rm C}-u_{\rm peak})/M=13.354$. At $a/M=0.99$, the first passing starts for the $4+2$ mode selection are $13M$ for strain and $2M$ for curvature, consistent with the spin-driven expansion of $\Delta v_{\mathrm{lr}\to r_0}$ to $15.096M$. At zero spin the curvature fit first passes at $1M$ (and the $2+2$ curve passes at $5M$ before merging with $4+2$ and $6+2$ near $u_S-u_{\rm peak}=9.435M$), while the strain fit has no persistent interval below $1\%$ in this scan.

\clearpage
\noindent\textit{Quantitative exterior onset benchmark table across spin.}
Table~\ref{tab:exterior-benchmark} lists the characteristic radii $(r_{\rm lr},r_+,r_0,r_-)$, the fundamental Kerr $(2,2,0)$ quasinormal frequency $M\omega_{220}$ obtained from Leaver's continued-fraction method, the post-light-ring intervals $\Delta v_{\mathrm{lr}\to +}$, $\Delta v_{\mathrm{lr}\to r_0}$, and $\Delta\tau_{\rm C}$ at $e_c=0$ and $e_c=0.8$, and the predicted exterior onset time $(u_{\rm C}-u_{\rm peak})/M$.

\begin{table}[!htbp]
\caption{\label{tab:exterior-benchmark}Exterior Kerr onset benchmark across prograde spin $a/M$, plus the $a/M=0.9$ FEW-matched eccentric plunge ($E=0.8447723$, $L_z/M=2.1020428$).}
\renewcommand{\arraystretch}{1.18}
\footnotesize
\begin{tabular}{@{}ccccccccccc@{}}
\toprule
$a/M$ & $r_{\rm lr}/M$ & $r_+/M$ & $r_0/M$ & $r_-/M$ & $M\omega_{220}$ & $\Delta v_{\mathrm{lr}\to +}^{(e_c=0)}$ & $\Delta v_{\mathrm{lr}\to r_0}^{(e_c=0)}$ & $\Delta\tau_{\rm C}^{(e_c=0)}$ & $\Delta\tau_{\rm C}^{(e_c=0.8)}$ & $(u_{\rm C}-u_{\rm peak})/M$ \\
\midrule
$0.00$ & $3.0000$ & $2.0000$ & $0.0000$ & $0.0000$ & $0.37367-0.08896i$ & $3.6462$ & $5.3861$ & $5.3861$ & $6.0652$ & $9.435$ \\
$0.10$ & $2.8822$ & $1.9950$ & $0.5903$ & $0.0050$ & $0.38823-0.08900i$ & $3.5693$ & $5.5274$ & $5.5457$ & $6.2561$ & $9.593$ \\
$0.20$ & $2.7592$ & $1.9798$ & $0.8110$ & $0.0202$ & $0.40459-0.08887i$ & $3.4868$ & $5.6265$ & $5.6663$ & $6.4108$ & $9.709$ \\
$0.35$ & $2.5627$ & $1.9367$ & $1.0222$ & $0.0633$ & $0.43360-0.08820i$ & $3.3528$ & $5.7510$ & $5.8593$ & $6.6634$ & $9.890$ \\
$0.50$ & $2.3473$ & $1.8660$ & $1.1538$ & $0.1340$ & $0.47036-0.08666i$ & $3.2077$ & $5.8815$ & $6.1281$ & $7.0189$ & $10.139$ \\
$0.60$ & $2.1889$ & $1.8000$ & $1.2082$ & $0.2000$ & $0.50153-0.08481i$ & $3.1069$ & $5.9981$ & $6.4001$ & $7.3766$ & $10.395$ \\
$0.70$ & $2.0133$ & $1.7141$ & $1.2369$ & $0.2859$ & $0.54125-0.08177i$ & $3.0073$ & $6.1781$ & $6.8241$ & $7.9268$ & $10.800$ \\
$0.80$ & $1.8111$ & $1.6000$ & $1.2367$ & $0.4000$ & $0.59560-0.07646i$ & $2.9221$ & $6.5170$ & $7.5835$ & $8.9001$ & $11.537$ \\
$0.90$ & $1.5579$ & $1.4359$ & $1.1952$ & $0.5641$ & $0.68143-0.06545i$ & $2.9135$ & $7.4056$ & $9.4095$ & $11.2216$ & $13.338$ \\
$0.95$ & $1.3863$ & $1.3122$ & $1.1450$ & $0.6878$ & $0.75532-0.05355i$ & $3.0660$ & $8.8023$ & $12.0955$ & $14.6872$ & $16.010$ \\
$0.99$ & $1.1676$ & $1.1411$ & $1.0628$ & $0.8589$ & $0.87671-0.02955i$ & $4.0817$ & $15.0962$ & $23.7117$ & $30.1450$ & $27.614$ \\
\midrule
$0.90$ (FEW) & $1.5579$ & $1.4359$ & $1.1956$ & $0.5641$ & $0.68143-0.06545i$ & $2.9254$ & $7.4196$ & $9.4253$ & --- & $13.354$ \\
\bottomrule
\end{tabular}
\end{table}

\medskip
\noindent\textit{Independent model and catalog checks.}
For the same masses and spin as the FEW example, we used an eccentric effective-one-body model with the same input eccentricity and initial average frequency~\cite{dali}. The amplitude peaks at $1151.11$ days from that model's initial conditions. The next sample is $48.3\%$ smaller at a separation of $1.48M_{\rm tot}$. A separate late-start calculation produces a substantially smaller adjacent-sample change. The EOB run follows its own orbit and merger attachment. The strain peak on the FEW-matched orbit requires its own perturbation calculation.

The six public Vishveshwara display traces at $q=M/\mu=12.1$~\cite{vwc} include the aligned eccentric case TEUK:BBH:00243 with $a/M=0.7$ and input eccentricity $0.2$. Its displayed amplitude first drops below half its peak at $16M_{\rm tot}$ after that peak. The comparable single-spin circular display crosses at $18M_{\rm tot}$. These displays include the real part and magnitude, and the holdout fit requires the full complex time series. This mass ratio is also far from the FEW baseline.

\end{document}